\RequirePackage{lineno}
\documentclass[aps,prc,twocolumn, floatfix]{revtex4}
\usepackage{graphicx}
\usepackage{xcolor}
\usepackage{dcolumn}
\usepackage{csquotes}
\usepackage{bm}
\usepackage{amsmath}
\usepackage{multirow}
\usepackage{float}

\begin{document}


\title{Constraining the Phase-Transition EoS using the Energy Dependence of Directed Flow}
\author{Zhi-Min Wu$^{1,2}$}
\author{Gao-Chan Yong$^{1,2,3}$}
\email[Corresponding author: ]{yonggaochan@impcas.ac.cn}
\author{Qingfeng Li$^{4,2}$}

\affiliation{
$^1$School of Nuclear Science and Technology, University of Chinese Academy of Sciences, Beijing 100049, China\\
$^2$Institute of Modern Physics, Chinese Academy of Sciences, Lanzhou 730000, China\\
$^3$State Key Laboratory of Heavy Ion Science and Technology, Institute of Modern Physics, Chinese Academy of Sciences, Lanzhou 730000, China\\
$^4$School of Science, Huzhou University, Huzhou 313000, China
}

\begin{abstract}

We propose a hybrid equation of state (VDF+MIT EoS) to describe the hadron-quark phase transition in dense nuclear matter. By coupling this EoS with the AMPT-HC transport model and comparing to recent experimental data on proton and $\Lambda$ directed flow $v_1$, we constrain the transition to likely occur near $5\rho_0$--$6\rho_0$, ruling out transitions below $3\rho_0$. Furthermore, we introduce the energy derivative of the mid-rapidity $v_1$ slope, $d(dv_1/dy)/d(\sqrt{s_{NN}})$, as a weakly model-dependent observable. Its zero crossing provides a direct signature of the phase transition critical point, offering a new tool for mapping the QCD phase diagram in future experiments.

\end{abstract}

\maketitle

\section{Introduction}
\label{sec:introduction}
The structure of the Quantum Chromodynamics (QCD) phase diagram, particularly its critical point and phase boundaries, constitutes a central problem in contemporary nuclear physics and astrophysics. On a macroscopic level, it shapes the thermodynamic and dynamical evolution of the early universe during its cooling phase \cite{OriginMatter},  whereas on a microscopic scale, it governs the confinement and deconfinement mechanisms of quarks and gluons that form baryonic matter \cite{QCDEoS}. Given the substantial unknown features within the QCD phase diagram, its full structure can only be unveiled through the study of physical systems at extremes of baryon density and temperature. Multimessenger astrophysical observations of compact stellar objects, including neutron stars, binary neutron star mergers, and supernovae, are indispensable in probing matter under such extreme conditions on cosmic scales \cite{GW1,NS1}. Complementarily, high-energy heavy-ion collision experiments serve as a highly controllable framework for recreating these extreme conditions in laboratory settings. By generating hot and dense matter, including the Quark-Gluon Plasma (QGP), such experiments offer unparalleled opportunities to systematically probe the properties of strongly interacting matter and verify predictions of the QCD phase diagram \cite{QCDphase}.

The commissioning of the Bevalac heavy-ion collider heralded the advent of relativistic heavy-ion physics, laying the groundwork for the first-ever laboratory creation of dense nuclear matter and the observation of anisotropic flow, a signature of collective motion within the collision system \cite{Bevalac}. Subsequently, high-energy experiments at next-generation accelerators like RHIC and the LHC established the existence of the QGP and provided critical evidence for its production, evolution, and jet quenching mechanisms \cite{exp1994}. However, to completely unravel the QCD phase diagram structure, particularly to search for the critical point and map the first-order phase boundary, the research focus has recently shifted toward the intermediate energy regime \cite{india,fxt2016}. The Beam Energy Scan (BES) program at RHIC-STAR addresses this by covering the phase transition sensitive energy region of 3.0-7.7 GeV \cite{highEnergy,star2019prc}, thereby filling a critical data gap in the high baryon density region and offering unprecedented opportunities for exploring the QCD phase structure.

The Equation of State (EoS) of nuclear matter governs the system dynamics during the thermalization stage of heavy-ion collisions, playing a pivotal role in nuclear physics and astrophysics \cite{nuclEoS, astroEoS}. It encapsulates fundamental thermodynamic properties that describe the extreme behavior of nuclear matter under varying densities and temperatures. Studies suggest that a QCD phase transition in dense matter may induce EoS softening \cite{softEoS,QCDpoint}, leading to intricate non-linear behaviors in regimes of high baryon density. However, deriving an accurate EoS at finite baryon chemical potential from first-principles calculations remains a formidable obstacle in nuclear physics. To address this, physicists often adopt transport models to simulate the spatiotemporal evolution of hot and dense nuclear matter \cite{yn2022prc, mok2022epjc}. Researchers constrain the EoS by calibrating model predictions against experimental data, thereby unveiling the system's dynamic properties under extreme conditions \cite{yn2017plb,manz2024,liqf2017,liuyy2021}. Recently, models incorporating a parameterized vector interaction within a relativistic density functional framework have successfully derived EoSs that not only comply with the known thermodynamic behavior of normal nuclear matter but also extend to the high baryon density region to include potential QCD phase transitions \cite{VDForigin}. Simulating the phase transition sensitive energy region by coupling such improved EoSs with transport models is expected to further shed light on the complex structure of the QCD phase diagram.

The directed flow $v_1$, which quantifies the sideward deflection of matter within the reaction plane, is a crucial dynamic observable whose characteristics are highly sensitive to the stiffness of the EoS of dense matter, making it a key probe for the EoS \cite{eventplanOrigin}. In this study, the AMPT-HC transport model, which employs a pure hadronic cascade, was utilized to simulate the phase transition-sensitive energy regime \cite{yong2021plb}. By embedding a hybrid EoS—constructed through a self-consistently formulated phenomenological hadronic EoS that incorporates vector-type interactions, coupled with a Quark EoS derived from the MIT bag model—into the transport framework, we investigated the influence of phase transition structures on baryon directed flow evolution. Integrating our findings with experimental data, we explored the critical behavior of the high-density nuclear EoS. Furthermore, we demonstrated that the rate of change of the directed flow slope $d(dv_1/dy)/d(\sqrt{s_{NN}})$ with respect to collision energy serves as a weakly model-dependent observable with extremely high sensitivity to phase transition signals. Quantitative measurement of this parameter can accurately describe the EoS's non-linear behavior at high densities and reveal the critical phase structure of dense nuclear matter. This observable is anticipated to provide new constraints for QCD phase diagram studies in future experiments, such as HIAF and FAIR.

\section{The AMPT-HC model with phase-transition equation of states}
\label{sec:methodology}
The AMPT (A Multi-Phase Transport) model, recognized for its comprehensive treatment of both partonic and hadronic phases and its ability to effectively capture non-equilibrium transport dynamics, has been widely applied in simulations of heavy-ion collisions at RHIC and LHC energy ranges \cite{nst2021}. As a Monte Carlo-based framework, AMPT integrates four fundamental components: initialization of initial-state fluctuations, partonic interactions, the conversion of partons into hadrons during hadronization, and subsequent hadronic interactions \cite{AMPT2005}. However, to explore the behavior of nuclear matter within the FXT energy range and the EoS that includes a ``QGP-like'' phase transition, the hadronic cascade mode of the AMPT model (referred to as AMPT-HC) is employed \cite{yong2021plb}. The AMPT-HC mode operates independently of partonic transport mechanisms. It uses the Woods-Saxon nuclear density profile in combination with the local Thomas-Fermi approximation to initialize the phase-space distribution of nucleons in both the projectile and target nuclei prior to collisions. Subsequently, it self-consistently describes the collision and transport dynamics of nucleons, baryon resonances, strange hadrons, and their corresponding antiparticles through a semi-classical test-particle method and hadronic mean-field potentials. Compared to the standard AMPT model, AMPT-HC is more suitable for studying intermediate-to-low-energy heavy-ion collisions, particularly in relation to collective motion dominated by hadronic dynamics. The operational principles and hadronic cascade dynamics of the AMPT-HC framework are detailed in several key references, including \cite{AMPT2005, deu2009, YongPRC2022, YongPRD2023, yong2024pt}.

\begin{table}[tp!]
    \centering
    \caption{Different constraint characteristics of QGP-like phase transition VDF EoSs, where the saturation density $\rho_0$, the binding energy $E_0$ are not shown. The incompressibility coefficients $K_0$ of the nuclear matter are obtained from the given EoS. A detailed description can be found in Refs. \cite{VDForigin,VDFlate}}
\label{tab:parameters}
\begin{ruledtabular}
      \begin{tabular}{@{} lcccccc @{}}
        \multirow{2}*{EoS}
        & {$T_{c}^{(N)}$}
        & {$\rho_{c}^{(Q)}$}
        & {$T_{c}^{(Q)}$}
        & {$\eta_{L}$}
        & {$\eta_{R}$}
        & {$K_0$ } \\
        & [MeV] & [$\rho_0$] & [MeV] & [$\rho_0$] & [$\rho_0$] & [MeV]  \\
    \hline
    VDF1  & 18    & 3.0   & 100   & 2.50  & 3.315  & 261   \\
    VDF2  & 18    & 4.0   & 50    & 3.85  & 4.124  & 279   \\
    VDF3  & 22    & 6.0   & 50    & 5.80  & 6.177  & 356   \\
      \end{tabular}
    \end{ruledtabular}
\end{table}

The relativistic Vector Density Functional (VDF) model, grounded in the theoretical framework of relativistic Landau Fermi liquid theory, adopts an effective quasi-particle approach whereby complex many-body interactions are encoded via a generalized energy density functional \cite{Landau1957,LFLmodel}. Through its flexible structure, the model incorporates higher-order density dependencies, enabling it to accurately reproduce the saturation properties of normal nuclear matter(e.g., saturation density $\rho_0 = 0.16 \, \text{fm}^{-3}$ and binding energy $E_0 = -16.3 \, \text{MeV}$). Moreover, it captures nontrivial phenomena, such as the nuclear liquid-gas phase transition and the QCD phase transition at high baryon density and temperature \cite{VDForigin}. This versatile EoS provides a solid theoretical foundation for exploring critical behavior related to high-density nuclear matter. Structurally similar to the Density Functional Theory (DFT) framework, the VDF model is capable of treating nuclear matter uniformly within the mean-field approximation. For application in transport models, the VDF framework naturally yields the single-particle potential suitable for the BUU/Boltzmann transport equation:
\begin{equation}\label{eq:1}
    U(\rho) = \sum_{i=1}^{N} \widetilde{C}_{i}(\rho/\rho_{0})^{b_i -1},
\end{equation}
here, $\rho$ denotes the baryon number density, and $b_i$ and $\widetilde{C}_i$ are VDF parameters. In this study, a parameterization incorporating four interaction terms was employed, yielding three distinct equations of state: VDF1, VDF2, and VDF3. The parameterization characteristics of these equations are summarized in Table~\ref{tab:parameters}, with detailed parameter sets provided in the references \cite{VDFlate}.

The original VDF model constructs its EoS based on quasi-particle degrees of freedom to characterize the QCD phase transition. However, the model lacks an effective description of the high-density quark matter properties in the post-transition region. Crucially, the strong repulsive interactions within the dense matter at high baryon number density cause the resulting EoS to violate the causality condition. To address this issue, we introduce the MIT bag model, thereby providing an enhanced characterization of the properties of quark matter.

In the absence of strange quarks and assuming the existence of only massless u and d quarks, the zero-temperature pressure $P_Q$ of quark matter in the MIT bag model is expressed as \cite{GYF2021}:
\begin{equation}\label{eq:2}
    \begin{split}
        P_Q = \frac{3}{4} \frac{\pi}{(\pi-2\alpha_S)^{1/3}}\rho^{4/3} - B,
    \end{split}
\end{equation}
where $B$ is the Bag Constant and the strong coupling constant $\alpha_s$ is set to $0.1$ \cite{mitbag1,mitbag2,mitbag3}. In contrast, the zero-temperature pressure $P_H$ of nuclear matter is typically composed of two contributions: the kinetic motion of particles and the mean-field potential interactions. For the widely applied Skyrme-type density-dependent interactions, the zero-temperature pressure can be written as:
\begin{equation}\label{eq:3}
    \begin{split}
        P_H =& g \int \frac{d^3 p}{(2\pi)^3}T \ln\left[ 1+ e^{-\beta(\varepsilon_p - \mu)}\right]\\
        &+\frac{\alpha}{2}\frac{\rho^2}{\rho_0} + \frac{\beta \gamma}{1+\gamma}\frac{\rho^{\gamma +1}}{\rho_{0}^{\gamma}},
    \end{split}
\end{equation}
where the first term reflects the contribution from the kinetic energy, while the latter two terms represent potential energy contributions. The Skyrme interaction parameters $\alpha, \beta, \text{ and } \gamma$ are typically determined by fitting ground state nuclear properties, such as the incompressibility coefficient $K_0$. The density dependence of $K_0$ can be parameterized via a polynomial form, which allows the Skyrme EoS to describe the transition properties connected to the quark phase. To achieve the desired transition between nuclear and quark matter, we adjust the Bag Constant $B$ such that the EoS derived from the VDF model smoothly transitions into the MIT EoS after the phase transition spinodal region. This extended EoS model provides a unified description encompassing the nuclear matter phase, the QCD phase transition region, and the quark matter phase, and has been successfully applied to transport model simulations. Detailed parameterization formulas and corresponding numerical optimization specifics are provided in Appendix ~\ref{app:a}.

For strange baryons ($\Lambda, \Sigma, \Xi, \text{ and } \Omega$), their interaction potential is grounded in the quark counting rule, which postulates that the interaction potential between strange baryons and non-strange baryons is proportional to the number of non-strange quarks within the baryon. The derivation and proof of this methodology can be found in the references ~\cite{hyperPoten1, hyperPoten2}. The K-meson mean-field potential utilizes parameters adopted from Ref. \cite{kaonPoten}, while the $\pi$-meson mean-field potential is, by default, not incorporated into the model calculation.

\section{Results and discussions}
\label{sec:result}
\begin{figure}[htp!]
    \centering
    \includegraphics[ width=0.45\textwidth]{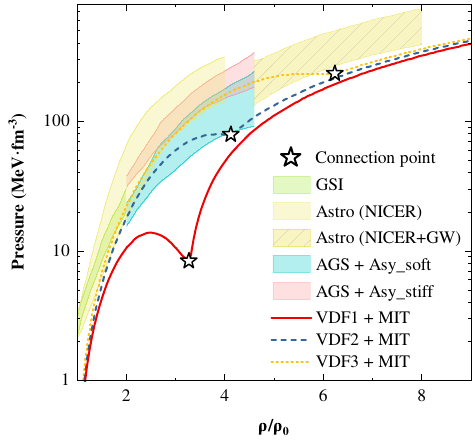}
    \caption{Pressure $P$ as a function of baryon number density $\rho$ for the three hybrid VDF+MIT EoSs. The calculated results for the three distinct EoSs are represented by a red solid line (VDF1+MIT), a blue dashed line (VDF2+MIT), and a yellow dotted-dashed line (VDF3+MIT). The hollow pentagrams indicate the transition points from the hadronic to the quark phase. The calculated EoSs are compared with constraints from astrophysical observations and experimental measurements (represented by the different color shaded band).} \label{Pressure}
\end{figure}

\begin{figure}
    \centering
    \includegraphics[width=0.45\textwidth]{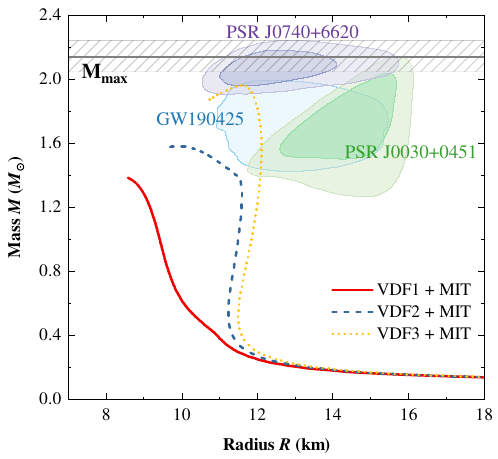}
    \caption{Mass-radius relations derived from the VDF+MIT EoS. Shaded regions represent observational constraints from PSR J0030+0451 (green) \cite{PSRJ0030p0451}, PSR J0740+6620 (purple) \cite{PSRJ0740p6620A,PSRJ0740p6620B}, and the GW190425 event (blue) \cite{NS1}. The grey band denotes the predicted theoretical maximum mass limit $M_{\max}$ \cite{maxM1,maxM2}.}\label{MRRelation}
\end{figure}

\begin{figure*}[ht!]
    \centering
    \includegraphics[width=0.9\textwidth]{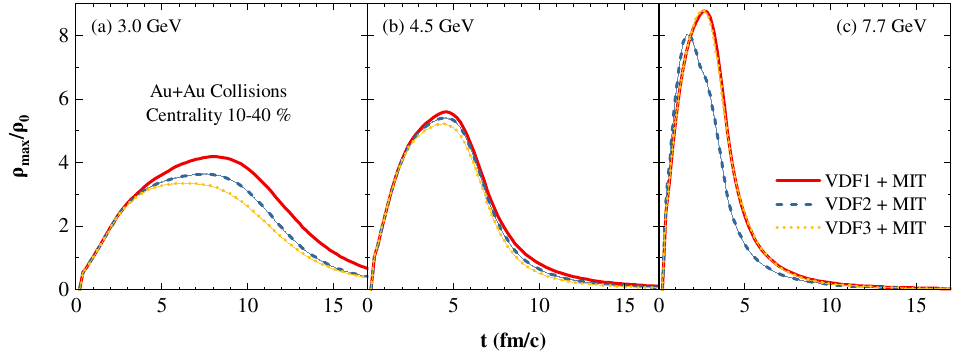}
    \caption{Time evolution of the maximum central compressional baryon number density ($\rho_{\text{max}}/\rho_0$) achieved in Au+Au collisions at various beam energies. The simulations were performed using the AMPT-HC transport model coupled with the VDF+MIT hybrid EoS.} \label{density}
\end{figure*}

Based on the methodology described in Section ~\ref{sec:methodology}, we successfully constructed a hybrid VDF+MIT EoS. This new EoS extends the original VDF model by allowing the hadronic phase to transition into the Quark EoS once the baryon density exceeds the spinodal region associated with the QCD phase transition. This crucial feature enables us to constrain the properties of dense matter beyond the phase boundary, providing a physically consistent description of high-density nuclear matter where the predicted speed of sound remains within acceptable limits. Figure ~\ref{Pressure} displays the pressure as a function of baryon number density for the newly constructed EoSs, alongside relevant constraints derived from astrophysical observations \cite{liaprd2023,apj2019} and heavy-ion collision experiments \cite{AGSsci2002,GSI2001,GSI2016npa,GSI2016prc}. From Figure ~\ref{Pressure}, one observes that EoS VDF1+MIT is inconsistent with both experimental and astrophysical constraints across the entire density range. In contrast, EoSs VDF2+MIT and VDF3+MIT are generally consistent with existing constraints. VDF3+MIT exhibits particularly strong agreement, with only minor deviations below twice the saturation density ($\rho \le 2\rho_0$) and above six times the saturation density ($\rho \ge 6\rho_0$). (The trend below $2\rho_0$ aligns with the AGS+$asy\_soft$ constraint, although this region is not the primary focus of the present study.)

The macroscopic structure of spherically symmetric, static neutron stars is governed by the Tolman-Oppenheimer-Volkoff (TOV) equations within the framework of general relativity \cite{tov1,tov2}. By numerically integrating these equations with our proposed hybrid VDF+MIT EoS, we derive the corresponding mass-radius ($M-R$) trajectories. As illustrated in Figure ~\ref{MRRelation}, the $M-R$ relationship exhibits a pronounced sensitivity to the structural evolution and the underlying stiffness of the EoS. Such sensitivity facilitates the use of multimessenger astrophysical observations—including binary neutron star mergers and pulsar timing—to impose stringent constraints on the nuclear EoS from a cosmological perspective. In Figure ~\ref{MRRelation}, the green and purple shaded regions represent the credible intervals for the pulsars PSR J0030+0451 \cite{PSRJ0030p0451} and PSR J0740+6620 \cite{PSRJ0740p6620A,PSRJ0740p6620B}, respectively, as characterized by NICER and XMM-Newton. The blue region denotes the constraints from the GW190425 event \cite{NS1}, while the grey hatched area indicates the current theoretical upper bounds for the maximum neutron star mass \cite{maxM1,maxM2}. Our demonstration reveals that the VDF3+MIT EoS demonstrates superior consistency with these observational benchmarks. Nevertheless, a marginal discrepancy persists between the calculated maximum mass and the theoretical limits. We posit that this residual tension may suggest the influence of dark matter components \cite{darkmatter1, darkmatter2} or potential deviations from general relativity in the strong-field regime \cite{NNG1, NNG2, NNG3}.

Furthermore, whether this phenomenon is intrinsically linked to the QCD phase transition critical point and the pressure evolution of quark matter remains a formidable challenge in high-density nuclear physics. As inferred from Figure ~\ref{Pressure} and ~\ref{MRRelation}, current astrophysical and laboratory constraints are insufficient to uniquely determine the occurrence of a phase transition in high-density matter. To provide more definitive insights into the hadron-quark transition regime, a more sensitive dynamical probe is required. In the following section, we incorporate the newly constructed EoS into the AMPT-HC model to simulate Au+Au collisions. By comparing the calculated directed flow of nucleons and hyperons with experimental data, we aim to extract concrete signatures of the hadron-quark phase transition.

\begin{figure*}[ht!]
    \centering
    \includegraphics[width=0.9\textwidth]{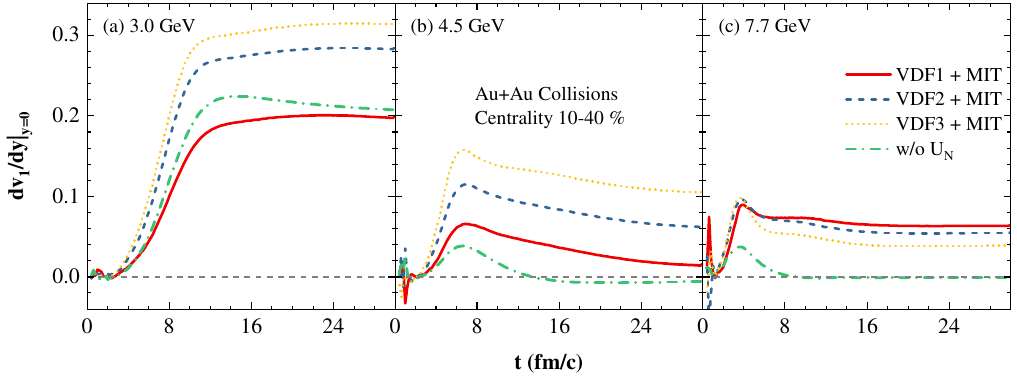}
    \caption{Time evolution of the proton directed flow $v_1$ slope at mid-rapidity in Au+Au collisions at $10-40\%$ centrality and center-of-mass energies of $\sqrt{s_{NN}} = 3, 4.5, \text{ and } 7.7 \, \text{GeV}$. The simulations were performed using the AMPT-HC transport model. Results using the three distinct VDF+MIT hybrid EoSs are shown by the red solid line (VDF1+MIT), the blue dashed line (VDF2+MIT), and the yellow dotted-dashed line (VDF3+MIT). The green dotted-dashed line represents the result from the pure hadronic cascade mode (no mean field).} \label{v1slope}
\end{figure*}

One of the key physical quantities related to the QCD phase diagram is the compressed baryon number density in heavy-ion collision experiments. Among these, heavy-ion collision experiments in the FXT energy region are especially noteworthy due to their high potential for probing the hadron-to-quark phase transition. Figure ~\ref{density} illustrates the evolution of the maximum compressed baryon number density as a function of collision energy and time, simulated using the AMPT-HC model combined with different phase transition EoSs in Au+Au collisions under FXT energies. From Figure ~\ref{density}, several observations can be made. Firstly, the maximum compressed baryon density generally increases with rising collision energy. However, stiffer (softer) EoSs, due to their strong repulsive interactions, make nuclear matter more difficult (easier) to compress, resulting in lower (higher) maximum compressed densities. Secondly, as collision energy increases, the relative differences in maximum compressed baryon density predicted by the VDF+MIT EoSs change gradually. At even higher energies, the maximum compressed density predicted by the VDF3+MIT EoS approaches that of the VDF1+MIT EoS. This convergence arises from the structural configuration of the VDF2+MIT and VDF3+MIT EoSs in high-density scenarios. Lastly, it can be observed that the VDF1+MIT EoS exceeds the phase transition threshold in Au+Au collisions at 3 GeV; the VDF2+MIT EoS exceeds the threshold at 4.5 GeV; and the VDF3+MIT EoS surpasses the threshold at 7.7 GeV. Therefore, the EoSs constructed using the VDF+MIT model should approximately delineate the baryon number density range for the hadron-to-quark phase transition in the FXT energy region, based on constraints from baryonic directed flow $v_1$ experimental data.

The directed flow $v_1$ is the first coefficient of the Fourier expansion of the final-state particle distribution in the azimuthal plane, typically defined as:
\begin{equation}\label{4}
    v_1 \equiv \langle \cos(\phi - \Psi_{RP}) \rangle = \langle \frac{p_x}{p_t} \rangle,
\end{equation}
where $\phi$ is the particle's azimuthal angle and $\Psi_{RP}$ is the reaction plane angle \cite{eventplanOrigin,eventplanRev}, which is conventionally set to $\Psi_{RP}=0$ in transport models. The magnitude of $v_1$ quantifies the preferential emission of particles along the direction of the reaction plane and is tightly coupled to the EoS of nuclear matter \cite{Bevalac,dv1,dv2,dv3,dv4,hs2000,hs1980,lp1999,pb2010}. Consequently, to investigate how the phase-transition EoS influences the formation and evolution of directed flow $v_1$, we simulated Au+Au collisions at three specific energy points within the FXT energy regime. The simulations utilized the AMPT-HC transport model coupled with the VDF+MIT hybrid EoS. We subsequently calculated the time evolution of the proton $v_1$ slope at mid-rapidity for the different phase-transition EoSs, as well as for the pure hadronic cascade mode (without mean field). As shown in Figure ~\ref{v1slope}, the proton $v_1$ mid-rapidity slope exhibits a strong sensitivity to the nuclear EoS. Furthermore, the introduction of the phase transition structure yields a unique spatiotemporal evolution pattern for $v_1$.

First, referencing the maximum baryon compression density evolution in Figure ~\ref{density}, we find that for the 3 GeV energy point, the $v_1$ slope primarily shows two stages: fast growth followed by slow growth. However, at the higher energies of 4.5 GeV and 7.7 GeV, the slope exhibits a pattern of initial rapid increase followed by a subsequent decay. Analysis of the pure HC mode calculations reveals that the observed $v_1$ decay primarily attributed to the re-scattering effect of particles emitted from the high-density zone by spectator nucleons \cite{Bevalac,dv1}. This effect is very weak in 3 GeV collisions, reaches its maximum suppression at 4.5 GeV, and then weakens at 7.7 GeV.

\begin{figure*}[ht!]
    \centering
    \includegraphics[width=0.9\textwidth]{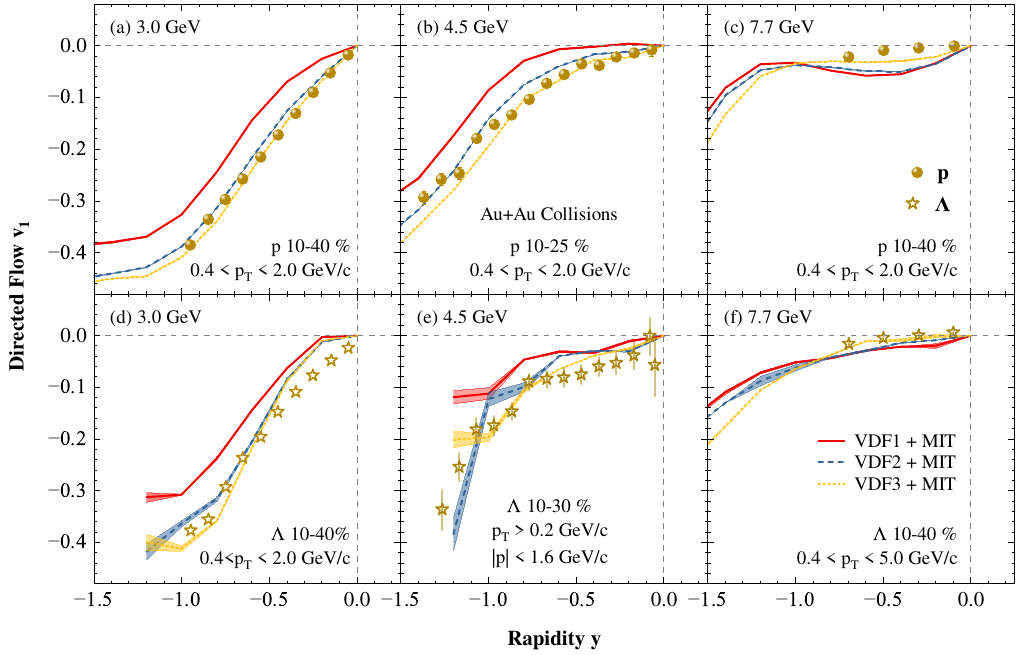}
    \caption{Directed flow ($v_1$) distribution as a function of rapidity ($y$) for protons and $\Lambda$ hyperons produced in Au+Au collisions. Model results corresponding to the three VDF+MIT hybrid EoSs are shown by the red solid line (VDF1+MIT), the blue dashed line (VDF2+MIT), and the yellow dotted-dashed line (VDF3+MIT), respectively. Experimental data for protons are represented by yellow circles and data for $\Lambda$ hyperons by yellow open stars. The experimental data points are sourced from the RHIC-STAR experiment in Au+Au collisions at center-of-mass energies of $\sqrt{s_{NN}} = 3.0, 4.5, \text{ and } 7.7 \, \text{GeV}$ \cite{starplb2022, star45, star77,star77p}.} \label{v1star}
\end{figure*}

Second, we observe a complex, energy-dependent influence of the mean field. At 3 GeV, the proton $v_1$ mid-rapidity slope obtained with the VDF1+MIT EoS is lower than that of the pure hadronic cascade. Conversely, at 7.7 GeV, the $v_1$ slopes from VDF2+MIT and VDF3+MIT are both weaker than the VDF1+MIT result. This behavior stems from the non-monotonic and unique structure of the phase-transition EoS within the mixed-phase region. This special structure generates a weak attractive potential as the system approaches the critical point during compression, slightly promoting compression. Conversely, it creates an inhibitory effect during the subsequent expansion phase as the system moves away from the critical point. For the 3 GeV collision, the VDF1+MIT EoS undergoes a phase transition at only $\sim 3\rho_0$ and features a relatively deep phase transition potential well. Since the maximum compression density at this energy is very close to the transition density, the phase transition structure effectively weakens the early lateral emission, thereby suppressing the proton $v_1$ formation, resulting in a value lower than the no-mean-field case. For the 7.7 GeV collision, the system's maximum compression density significantly exceeds the transition densities of VDF2+MIT ($\sim 4\rho_0$) and VDF3+MIT ($\sim 6\rho_0$). Crucially, the system spends a considerable amount of time in its respective mixed-phase region during the lengthy expansion phase at this high energy. This prolonged stay in the mixed phase results in a deficiency of transverse pressure during expansion. Consequently, the VDF3+MIT EoS, which spends the most time in this region, exhibits the fastest decline in $v_1$ strength, followed by VDF2+MIT EoS, leading to the weakest $v_1$ at the final state.

In summary, our study reveals that the presence of a phase transition structure in the EoS leads to a significant suppression of the $v_1$ slope. Crucially, this effect on the particle transport process is most pronounced when the system's maximum compressional density closely approaches the phase transition region. Consequently, a systematic analysis of this phenomenon in experimental directed flow data will be instrumental in deepening our understanding of the relationship between the high-density nuclear EoS and baryon number density. Ultimately, this analysis will provide a critical and sensitive phenomenological constraint toward the definitive determination of the QCD phase diagram.

\begin{figure*}[ht!]
    \centering
    \includegraphics[width=0.9\textwidth]{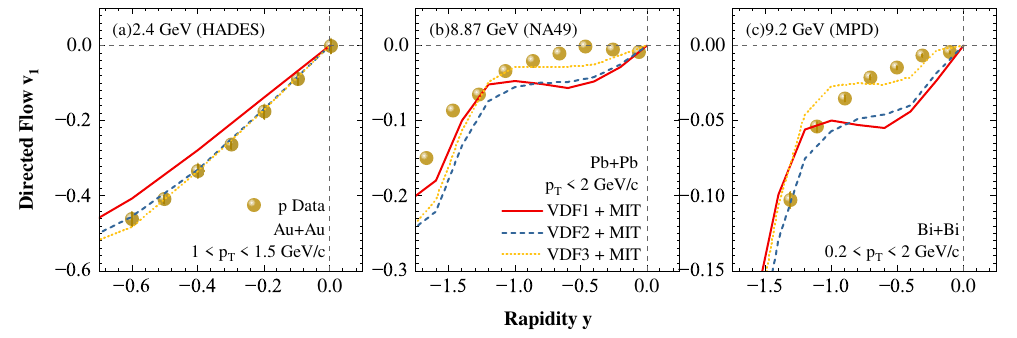}
    \caption{Directed flow $v_1$ distribution of protons as a function of rapidity $y$ from various experimental sources compared with model calculations. In all panels, the experimental data are represented by yellow circles. The lines correspond to the calculated results using the three distinct VDF+MIT hybrid EoSs. The panels show: (left) HADES $Au+Au$ collisions at $\sqrt{s_{NN}} = 2.4$ GeV \cite{hades2020}; (middle) NA49 $Pb+Pb$ collisions at $p_{beam} = 40$ GeV/c \cite{NA49}; and (right) MPD $Bi+Bi$ collisions at $\sqrt{s_{NN}} = 9.2$ GeV \cite{mpd2025}.}
    \label{hadesToMPD}
\end{figure*}

The RHIC-STAR BES program aims to explore the QCD phase transition critical point and phase boundaries, as well as the EoS of high-density nuclear matter. The recent measurement of the directed flow $v_1$ distribution as a function of rapidity $y$ for protons and $\Lambda$ hyperons at three energy points provides a crucial experimental benchmark for theoretical models. Experimental data suggest an EoS characterized by a stiff behavior at low densities that subsequently softens at high densities. The AMPT-HC transport model, coupled with the VDF+MIT EoS incorporating a phase transition structure, effectively captures this characteristic. As illustrated in Figure ~\ref{v1star}, the experimental data for both proton and $\Lambda$ hyperon $v_1$ distributions strongly favor the VDF3+MIT EoS. This EoS describes a scenario where nuclear matter is stiff during initial compression and undergoes softening at densities between $5\rho_0$ and $6\rho_0$. While the VDF2+MIT EoS (with a phase transition at $\sim 4\rho_0$ to $5\rho_0$) cannot be entirely ruled out, the VDF1+MIT EoS (with a transition at $\sim 3\rho_0$) is demonstrably unable to reproduce the measured $v_1(y)$ distribution within its applicable energy range. Therefore, our $v_1$ analysis provides strong confidence to exclude EoSs that predict a phase transition occurring below $3\rho_0$ and supports the conclusion that the hadron-quark phase transition is likely located near $5\rho_0$ to $6\rho_0$.

Complementary experimental results from various collaborations provide critical cross-verification for our theoretical framework. The HADES data at $\sqrt{s_{NN}} = 2.4$ GeV (Au+Au system) \cite{hades2020} probe the evolution at relatively lower baryonic compression densities. Conversely, the NA49 experiment at $p_{beam} = 40$ GeV/c ($\sqrt{s_{NN}} \approx 8.87$ GeV) \cite{NA49} and the recent inaugural MPD run at $\sqrt{s_{NN}} = 9.2$ GeV (Bi+Bi system) \cite{mpd2025} extend our investigation to regimes of higher baryon density. As illustrated in the three panels of Figure ~\ref{hadesToMPD}, the measured proton directed flow $v_1$ across these energies exhibits a consistent trend with the RHIC-STAR results at 3.0 and 7.7 GeV. These observations collectively favor an EoS incorporating a phase transition structure within the $5\text{–}6\rho_0$ range, reinforcing the necessity of quark-matter contributions beyond the transition threshold. This systematic analysis across a wide density span suggests that the hadron-quark transition likely occurs in the high-density region of $5\text{–}6\rho_0$. Consequently, future high-statistics and high-precision beam energy scans utilizing advanced detection technologies will be indispensable for imposing rigorous constraints on the high-density EoS, the QCD critical point, and the phase boundary.

Relying on experimental data from only a limited number of energy points can merely offer preliminary validation for the existence of a phase-transition EoS and provide a rough estimate for the critical point's location. Consequently, identifying observables that are highly sensitive to the critical point using available data remains a persistent research challenge. Looking ahead, the situation is evolving rapidly: on one hand, RHIC-STAR is scheduled to perform measurements at additional energy points within the FXT energy regime ($3.0-7.7 \, \text{GeV}$); on the other hand, forthcoming experiments, such as GSI-CBM and HIAF-CEE \cite{cbm,hades,mpd,hiaf,jap}, are poised to conduct dedicated studies in the phase transition sensitive region. The rich volume of future experimental data will provide reliable support for theoretical investigations of the QCD phase transition. However, the pressing concern remains how to efficiently and effectively extract information regarding the phase transition critical point from this wealth of data.

\begin{figure*}[ht!]
    \centering
    \includegraphics[width=0.9\textwidth]{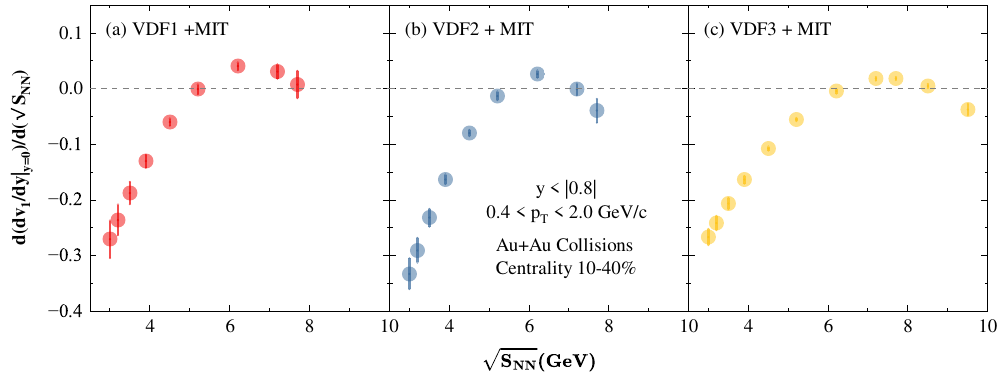}
    \caption{Proton $d(d v_1/dy )/d(\sqrt{s_{NN}})$ as a function of $\sqrt{s_{NN}}$ for the three VDF+MIT EoSs.}
    \label{rate}
\end{figure*}

As previously established, the directed flow $v_1$ strength experiences a more pronounced suppression due to the existence of the phase transition region when the maximum compressional baryon density achieved in the heavy-ion collision is closer to the phase transition critical point. Since reactions at different beam energies correspond to distinct spatiotemporal density evolutions, analyzing the change in the mid-rapidity $v_1$ slope as a function of the center-of-mass energy $\sqrt{s_{NN}}$ allows us to indirectly infer whether the maximum compression density at that energy is approaching the critical point. Therefore, we performed a transport model simulation scan across several energy points in the FXT regime and at higher energies, analyzing the proton mid-rapidity $v_1$ slope. Based on the distribution of the proton mid-rapidity slope versus $\sqrt{s_{NN}}$, we obtained a fitted curve using polynomial regression. Subsequently, we derived the rate of change of the mid-rapidity $v_1$ slope with respect to $\sqrt{s_{NN}}$ by taking the derivative of the fitted polynomial.

As demonstrated in Figure ~\ref{rate}, the calculated rate of change of the proton mid-rapidity $v_1$ slope, $d(d v_1/dy )/d(\sqrt{s_{NN}})$ for all three EoSs exhibits a similar non-monotonic evolution trend. Specifically, the rate of change transitions from a negative value to a positive value as energy increases, before dropping back into the negative range at even higher energies. This trend can be attributed to the interplay of three factors: First, as the collision energy increases, the maximum compressional density approaches the phase transition region. The suppression of $v_1$ due to the EoS's phase transition structure reaches its strongest point, leading to the first zero crossing of the rate of change. Second, with a continued increase in energy, the system rapidly passes through the phase transition region. The phase structure's suppressive effect weakens, and the repulsive mean field from nucleon-nucleon interactions enhances the $v_1$ formation, keeping the slope's rate of change positive.
Third, at higher energies, the contribution of the mean field to $v_1$ gradually saturates. Simultaneously, the baryonic re-scattering effect caused by spectator nuclear matter continues to suppress the transverse emission. This combined effect causes the overall $v_1$ strength to decrease with energy, pulling the rate of change back into the negative range. Crucially, the VDF1+MIT EoS (with a transition at $3\rho_0$) reaches this zero point at a significantly lower energy, while the VDF3+MIT EoS (requiring a transition at $5\rho_0 \sim 6\rho_0$) necessitates a higher collision energy to reach its zero point. This zero point serves as a distinct signal that the high-density matter produced in the collision has crossed the phase transition critical point and is largely situated within the mixed-phase/spinodal region. Precise experimental measurement of this QCD phase transition signal would greatly advance our understanding of the phase structure of the high-density nuclear EoS and provide an effective constraint for locating the QCD critical point and phase boundaries.

\section{Conclusions}
\label{sec:conclusion}
We extended the Relativistic Vector Density Functional (VDF) EoS by incorporating the MIT Bag Model to better describe high-density nuclear matter properties and the phase transition structure. Utilizing this improved EoS, coupled with the AMPT-HC transport model, we calculated the directed flow $v_1$ for heavy-ion collisions in the phase transition sensitive energy region. By comparing our simulations with experimental data, we obtained tighter constraints on the high-density nuclear EoS and the location of the QCD phase transition critical point. Furthermore, we propose a new weakly model-dependent observable that quantifies the stiffness of the high-density EoS as a function of the maximum compressional density. Experimental identification of the ``zero crossing'' point of this observable will serve as a crucial tool for precisely locating the QCD phase transition.

\section{Acknowledgments}
\label{sec:acknowledgment}
This work is supported by the National Natural Science Foundation of China under Grant Nos. 12275322, 12335008 and CAS Project for Young Scientists in Basic Research YSBR-088.

\appendix
\section{Detailed Parametrization Process of VDF+MIT EoS}\label{app:a}

Analogous to the zero-temperature nuclear matter pressure provided by Skyrme-type interactions Eq. \eqref{eq:3}, the zero-temperature pressure $P(\rho)$ of the VDF model in the rest frame can be expressed as:
\begin{equation}\label{VDFpressure}
    \begin{split}
        P(\rho)\bigg|_{\substack{\text{rest} \\ \text{frame}}} \approx& g \int \frac{d^3 p}{(2\pi)^3}T \ln\left[ 1+ e^{-\beta(\varepsilon_p - \mu)}\right]\\
        &+ \sum_{i=1}^{N=4} \frac{\widetilde{C}_i}{b_i}(b_i-1)\frac{\rho^{b_i}}{\rho_{0}^{b_i-1}}.
    \end{split}
\end{equation}
Here, $T$ is the temperature, $\beta = 1/T$, $\varepsilon_p$ is the particle energy, $\mu$ is the chemical potential, and the degeneracy factor is set to $g=4$. In the zero-temperature limit for an ideal Fermi gas, the kinetic pressure ($P_{\text{kin}}$), derived from the first term of Eq. \eqref{VDFpressure} (ideal Fermi gas approximation), simplifies to:
\begin{equation}\label{kinP0}
    \begin{split}
        P_{\text{kin}}(\rho) &= \frac{g}{2 \pi^2}\int_{0}^{p_F} \frac{p^4}{3E}dp\\
        &=\frac{g}{16 \pi^2}\left[\frac{2}{3}E_F p_F^3 - m^2E_F p_F +m^4 \ln \left(\frac{E_F + p_F}{m}\right) \right],
    \end{split}
\end{equation}
where the Fermi momentum $p_F$ and Fermi energy $E_F$ are defined, respectively, by the relations $p_F = \hbar \left( \frac{3\pi^2 \rho}{2} \right)^{1/3}$ and $E_F = \sqrt{p_F^2 + m^2}$.

The Skyrme parameters $\alpha, \beta$, and $\gamma$ can be determined by the incompressibility coefficient $K_0$, through the following relations:
\begin{align}
    \label{alpha}
        \alpha &= -29.81 - 46.9(K_0 + 44.73)/(K_0 - 166.32), \\
    \label{beta}
        \beta &= 23.45 (K_0 + 255.78)/(K_0 - 166.32), \\
    \label{gamma}
        \gamma &= (K_0 + 44.73)/211.05.
\end{align}

We first tune the Bag Constant $B$ such that the nuclear matter pressure given by the VDF EoS (Eq. \eqref{VDFpressure}) transitions into the quark matter pressure given by the MIT Bag Model (Eq. \eqref{eq:2}) after crossing the spinodal region of the phase transition. Since the incompressibility coefficient $K_0$ can be expanded into a density-dependent polynomial form $K_0(\rho/\rho_0)$, we fit the MIT Bag Model pressure using the Skyrme EoS pressure (Eq. \eqref{eq:3}) along with the $K_0$ relations (Eqs. \eqref{alpha}-\eqref{gamma}) to derive the polynomial expression for $K_0(\rho/\rho_0)$. The corresponding Bag Constant $B$ and the polynomial expression for $K_0(\rho/\rho_0)$ for the three VDF EoSs are summarized below:
For VDF1, $B = 130 \text{ MeV/fm}^3$, and $K_0(\rho/\rho_0)$ as:
\begin{equation}\label{VDF1MIT}
    \begin{split}
      K_0(\rho/\rho_0) = & 0.415(\rho/\rho_0)^3 - 10.422(\rho/\rho_0)^2 + \\
      &80.66(\rho/\rho_0) - 44.73.
    \end{split}
\end{equation}
For VDF2, $B = 108 \text{ MeV/fm}^3$, and
\begin{equation}\label{VDF2MIT}
    \begin{split}
      K_0(\rho/\rho_0) = & 0.588(\rho/\rho_0)^3 - 13.387(\rho/\rho_0)^2 + \\
      &91.682(\rho/\rho_0) - 26.581.
    \end{split}
\end{equation}
For VDF3, $B = 90 \text{ MeV/fm}^3$, and
\begin{equation}\label{VDF3MIT}
    \begin{split}
      K_0(\rho/\rho_0) = & 0.3396(\rho/\rho_0)^3 - 7.217(\rho/\rho_0)^2 + \\
      &40.417(\rho/\rho_0) + 119.266.
    \end{split}
\end{equation}
Therefore, we utilize the aforementioned approximation method to implement the hybrid VDF+MIT EoS in the transport model simulations.

\end{document}